\documentclass[aps,prl,twocolumn,showpacs,groupedaddress]{revtex4}
\usepackage{graphicx}  
\usepackage{dcolumn}   
\usepackage{bm}        
\usepackage{amssymb}   

\begin{document}

\title{Non-Newtonian fluid flow through three-dimensional disordered porous 
media}

\author{Apiano F. Morais$^{1}$, Hansjoerg Seybold$^{2}$, 
Hans J. Herrmann$^{1,2}$ and Jos\'e S.~Andrade Jr.$^{1,2}$}

\affiliation{$^{1}$Departamento de F\'{\i}sica, Universidade Federal 
do Cear\'a, 60451-970 Fortaleza, Cear\'a, Brazil\\ $^{2}$Computationla Physics, IfB, Schafmattst. 6, ETH,
8093 Zurich, Switzerland}

\date{\today}

\begin{abstract}
We investigate the flow of various non-Newtonian fluids through
three-dimensional disordered porous media by direct numerical
simulation of momentum transport and continuity equations. Remarkably,
our results for power-law (PL) fluids indicate that the flow, when
quantified in terms of a properly modified permeability-like index and
Reynolds number, can be successfully described by a single (universal)
curve over a broad range of Reynolds conditions and power-law
exponents. We also study the flow behavior of Bingham fluids described
in terms of the Herschel-Bulkley model. In this case, our simulations
reveal that the interplay of ({\it i}) the disordered geometry of the
pore space, ({\it ii}) the fluid rheological properties, and ({\it
iii}) the inertial effects on the flow is responsible for a
substantial enhancement of the macroscopic hydraulic conductance of
the system at intermediate Reynolds conditions. This anomalous
condition of ``enhanced transport'' represents a novel feature for
flow in porous materials.
\end{abstract}


\maketitle

The research on flow through porous media has great relevance for many
problems of practical interest in several fields, including physics,
medicine, biology, chemical and mechanical engineering and geology
\cite{Dullien79,Adler92,Sahimi95}. The disordered aspect of most natural 
and artificial porous materials is directly responsible for the
presence of local flow heterogeneities that can dramatically affect
the behavior, for example, of the transport of heat and mass through
the system. Under this framework, the standard approach to investigate
single-phase flow in porous media is to apply Darcy's law
\cite{Dullien79,Adler92,Sahimi95}, which simply assumes that a global
permeability $k$ relates the average fluid velocity $u_0$ in the
field with the pressure drop $\Delta p$ measured across the system,
\begin{equation}
\label{eq:Darcy}
u_0 = -{k \over \mu}{\Delta p \over L},
\end{equation}
where $L$ is the length of the sample in flow direction and $\mu$ is
the viscosity of the fluid. As a macroscopic index, the permeability
reflects the relation between the complex pore space morphology and
fluid flow. 

In previous studies
\cite{Canceliere90,Kostek92,Martys94,Andrade95,Koponen97,Rojas98,Andrade99},
detailed models of pore geometry have been used in combination with
computational fluid dynamics simulations to predict permeability
coefficients and validate classical semi-empirical correlations for
real porous materials.
\begin{figure}
\includegraphics[width=8cm]{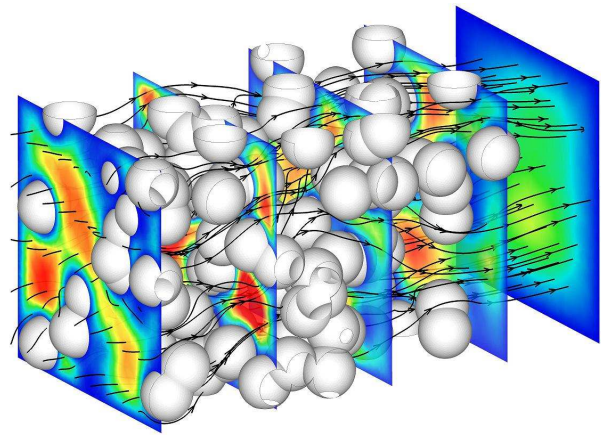}
\caption{(Color online) Non-Newtonian (power-law) fluid flow through 
a typical realization of the Swiss-Cheese pore space ($\varepsilon=0.7$). 
The fluid is pushed from left to right. The solid lines with arrows
correspond to trajectories of tracer particles released in the flow,
while the contour plots give the velocity magnitude at different
cross-sections of the porous medium. Their colors ranging from blue
(dark) to red (light) correspond to low and high velocity magnitudes,
respectively.}\label{fig1}
\end{figure}
In principle, the original concept of permeability as a global index
for flow in porous media, however, is only applicable in the limit of
Stokesian flow (linear). Strictly speaking, the validity of Darcy's
law should be restricted to (\textit{i}) Newtonian fluids and
(\textit{ii}) flows under viscous conditions, i.e., flows at very low
Reynolds numbers, defined usually as $\mathrm{Re} \equiv \rho u_0 d_p
/\mu$, where $\rho$ is the density of the fluid and $d_p$ is the grain
diameter. The departure from Darcy's law due to the contribution of
inertial forces (convection) to the flow of Newtonian fluids has been
the subject of several studies in the past
\cite{Edwards90,Koch97,Andrade99}. In particular, it has been
experimentally and numerically observed that the breakdown of
condition (\textit{ii}) can take place even under laminar flow
conditions, i.e., before fully developed turbulence effects become
relevant to momentum transport.

In order to understand the physics of important problems like blood
flow through the kidney \cite{Mattson93} or oil flow through porous rocks
\cite{McCain90}, for example, one has to overcome the restriction (\textit{i})
mentioned above by explicitly considering the nonlinear behavior of
these fluids under shear, namely, their specific non-Newtonian
properties. Although these fluids have been known for a long time,
technological applications which directly make use of their anomalous
rheological behavior have come into focus only recently. For instance,
shear thinning solvents are present in dropless paints
\cite{Maestro02}, and shear thickening fluids are currently used as
active dampers and components of enhanced body armors
\cite{Decker07}. While the physical properties of Newtonian fluid flow
through irregular media are theoretically well understood and have
been confirmed by many experiments, non-Newtonian systems
\cite{Metzner56,Sahimi93,Chhabra01} lack a generalized description. In this
Letter we investigate the flow of non-Newtonian fluids through
three-dimensional porous media by direct numerical simulation of
momentum and continuity equations. To the best of our knowledge, this
is the first time that nonlinear effects coming from both rheological
and inertial aspects of the fluid flow are considered simultaneously
in the framework of a disordered three-dimensional pore space.

The porous medium studied here is a three-dimensional realization of
the Swiss-Cheese model \cite{Lorenz01}. Spherical particles (solid
obstacles) of diameter $d_p$ are sequentially and randomly placed
in a box of length $L$ in the $x$-direction and square cross-section
of area $A$. Particle overlap is allowed and the allocation process
continues up to the point in which a prescribed value for the porosity
(void fraction) $\varepsilon$ is achieved. The mathematical formulation
for the fluid mechanics in the interstitial pore space is based on the
assumptions that we have a continuum and incompressible fluid flowing
under steady-state and isothermal conditions. Thus, the momentum and
mass conservation equations reduce to,
\begin{eqnarray}
\label{eq:MomentumNS}
\rho\vec{u}\cdot\nabla\vec{u}&=&-\nabla p+\nabla \mathcal{T}\\
\nabla\cdot\vec{u}&=&0~,
\end{eqnarray}
where $\vec{u}$ and $p$ are the velocity and pressure fields,
respectively, and $\mathcal{T}$ is the so-called deviatoric stress
tensor given by,
\begin{equation}
\label{eq:DeviatoricStress}
\mathcal{T}_{ij}=2\mu\dot{s}_{ij}~,
\end{equation}
where $\dot{s}_{ij}=\frac{1}{2}\left(\frac{\partial u_i}{\partial
x_j}+\frac{\partial u_j}{\partial x_i}\right)$ is the strain rate
tensor. The variable $\mu$ is the dynamic viscosity, for which a
constitutive relation must be provided in order to describe the
specific non-Newtonian behavior of the fluid. Here we investigate the
flow of two different types of rheologies, namely, the cross-power-law
fluid and the Bingham fluid. The constitutive relation for a PL fluid
can be written as,
\begin{equation}
\label{eq:PowerLawFluid}
\mu=K\dot{\gamma}^{n-1}, \qquad  \mu_1 < \mu < \mu_2,
\end{equation}
where the constants $\mu_1$ and $\mu_2$ are the lower and upper cutoffs,
$\dot{\gamma} \equiv \sqrt{\frac{1}{2} \dot{s}_{ij}
\dot{s}_{ij}}$ is the effective strain rate, $K$ is 
the consistency index, and $n$ is the power-law exponent. For $n=1$ we
recover the behavior of a Newtonian fluid. Fluids with $n>1$ are
shear-thickening, while shear-thinning behavior corresponds to $n<1$.

In the case of Bingham fluids, the rheology is commonly approximated
by the Herschel-Bulkley model \cite{Herschel26,Tang04} which combines
the effects of Bingham and power-law behavior for a fluid. For low
strain rates, $\dot{\gamma} < \tau_0/\mu_0$, the material acts as a
very viscous fluid with viscosity $\mu_0$. As the strain rate
increases and the yield stress threshold $\tau_0$ is surpassed, the
fluid behavior is described by,
\begin{equation}
  \label{eq:Bingham}
\mu=\frac{\tau_0+K_{B}[\dot{\gamma}^{n}-(\tau_0/\mu_0)^{n}]}
{\dot{\gamma}},
\end{equation}
where $K_{B}$ is the consistency factor and $n$ is the power-law
index. Here we restrict our simulations to the case of Bingham fluids
$n=1$, i.e., the fluid is still Newtonian at large strain rates, with 
a viscosity $\mu=K_{B}$.

Non-slip boundary conditions are applied along the entire solid-fluid
interface and end effects on the flow field, which become significant
at high Reynolds numbers, are minimized by attaching ancillary zones
at the inlet and outlet of the two opposite faces in the direction of
the flow (i.e., $x$-direction). At the inlet, a constant inflow
velocity in the normal direction to the boundary is specified, whereas
at the outlet we impose gradientless boundary condition. Finally, the
four remaining faces are considered to be solid walls.

For a given realization of the porous medium and a given set of flow
and constitutive parameters of the fluid, the numerical solution of
the partial differential equations (\ref{eq:MomentumNS}) for the local
velocity and pressure fields in the fluid phase of the void space,
head and recovery zones is obtained by discretization using the
control volume finite-difference technique \cite{Patankar80}.  An
unstructured grid with up to three million tetrahedral cells is
adapted to the geometry of the porous medium. For comparison, entirely
consistent numerical solutions have also been calculated with a
finite-volume scheme \cite{OpenFoam}. Finally, from the area-averaged
pressures at the inlet and outlet positions, the overall pressure drop
can be readily calculated.

\begin{figure}
\includegraphics[width=8cm]{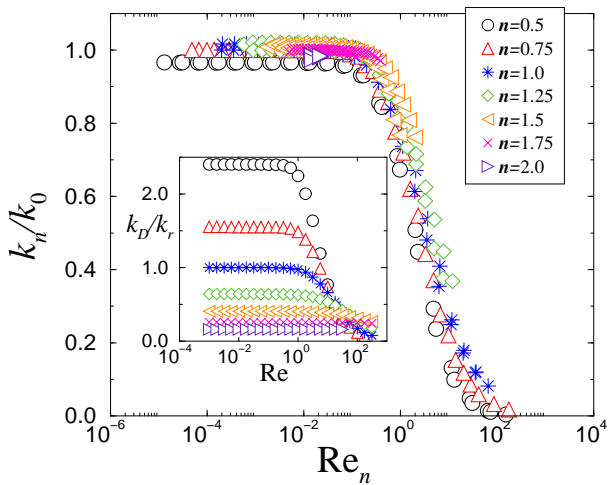}\caption{(Color online) Flow of 
power-law fluids through three-dimensional porous media. The inset
shows the variation of the ratio $k_{D}/k_{r}$ with Reynolds number
$\mathrm{Re} \equiv \rho u_0 d_p /K_{1}$ for different values of the
power-law exponent $n$ and $\varepsilon=0.5$. The resulting data
collapse presented in the main plot confirms the adequacy of our
rescaling procedure in terms of the modified permeability index
$k_{n}/k_{0}$ and the modified Reynolds number $\mathrm{Re}_{n}$ (see
text).}
\label{fig2}
\end{figure}

In Fig.~\ref{fig1} we show a three-dimensional plot of a typical
realization of the porous medium through which a power-law fluid
flows. Clearly, the complex geometry of the pore space induces
preferential channels on the flow whose localization and strength are
significantly dependent on the rheological properties of the fluid as
well as on the imposed inlet-outlet boundary conditions. For PL
fluids, this intricate interplay between geometry and flow can
nevertheless be macroscopically quantified in terms of an analogous to
a permeability index, namely a hydraulic conductivity, defined in
terms of Darcy's law as $k_{D} \equiv K_{1}u_{0}L/\Delta p$, where
$K_{1}$ is a reference viscosity taken as the consistency index for
$n=1$. As shown in the inset of Fig.~\ref{fig2}, the general behavior
of $k_{D}$ is qualitatively similar for different values of the
exponent $n$. Moreover, it follows the characteristic trend of a
simple Newtonian fluid ($n=1$), namely that $k_{D}$ remains
essentially invariant for low $\mathrm{Re}$ values up to a crossover
point $\mathrm{Re}_{\times}$ where it starts to decrease due to the
onset of non-linear convective effects on the flow
\cite{Andrade95,Andrade99,Andrade07}. Quantitatively, however, we
observe that both the upper limit for $k_D$ and $\mathrm{Re}_{\times}$
are strongly dependent on $n$.

Darcy's law has been generalized to power-law fluids in previous 
studies \cite{Bird07,Larson81,Yortsos95}. Here we define an hydraulic
conductivity as,
\begin{equation}
\label{eq:mod_perm}
k_{n} \equiv u_{0}K\left(\frac{\Delta p}{L}\right)^{-1/n}.
\end{equation}
\begin{figure}
\includegraphics[width=8cm]{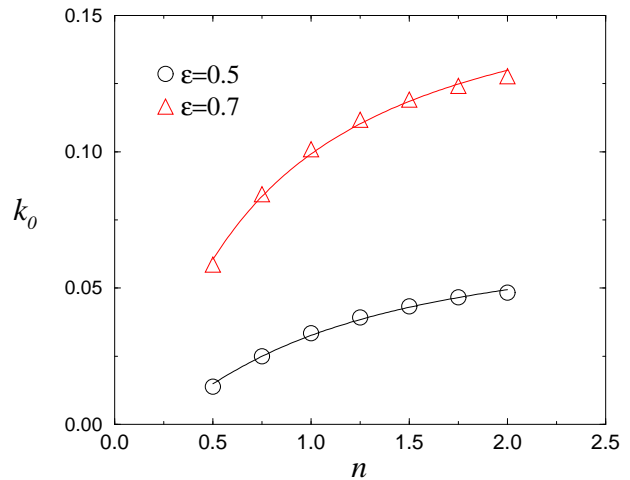}
\caption{(Color online) Dependence of the hydraulic conductivity at 
very low Reynolds numbers $k_{0}$ on the power-law exponent $n$ for
two different values of porosity $\varepsilon$. The solid lines are
the least-squares fits to the simulation data using Eq.~(\ref{eq:Bird}) 
with $d_{e}/d_{p}=0.35$ and $1.58$, for $\varepsilon=0.5$ and $0.7$, 
respectively.}
\label{fig3}
\end{figure}
As shown in Fig.~\ref{fig3}, this generalized index when calculated at
low Reynolds numbers, namely $k_{0}$, can be consistently correlated
with intrinsic properties of the fluid and porous medium by means of
the following semi-empirical expression
\cite{Bird07,Yortsos95}:
\begin{equation}\label{eq:Bird}
k_{0} = \frac{12}{25}\frac{n\left(75k_{r}\right)^{1/n}}{3n+1}
\frac{d_{e}^{\left(n-1\right)/n}}{3^{\left(n+1\right)/n}}\varepsilon^{2\left(1-n\right)/n}
K^{\left(n-1\right)/n}~,
\end{equation} 
where $d_{e}$ is the only fitting parameter corresponding to an
average effective pore diameter, namely, the average pore size (in
units of $d_p$) of the system calculated as if it was a packed bed
consisting of identical spheres \cite{Bird07,Christopher65}. The 
parameter $k_{r}$ corresponds to the value of $k_{D}$ calculated for a
Newtonian fluid ($n=1$) under very low Reynolds conditions, i.e., the
porous medium permeability according to Darcy's law.

In order to substantiate the non-Newtonian aspect of the fluid, it is also 
necessary to redefine the Reynolds number as \cite{Christopher65},
\begin{equation}
\label{reynolds}
\mathrm{Re}_{n}\equiv\frac{k_{0}^n\rho u_0^{2-n}}{2K^{n} d_p}\frac{1-\varepsilon}{\varepsilon^3}~,
\end{equation}
where the term $(1-\varepsilon)/\varepsilon^{3}$ has been adapted from
the classical Kozeny-Carman equation \cite{Dullien79}. It is worth
mentioning that Eq.~(\ref{reynolds}) breaks down close to the critical
percolation porosity \cite{Sahimi94}.

In the main plot of Fig.~\ref{fig2}, we show that all data sets of
$k_{n}/k_{0}$ against $\mathrm{Re}_{n}$, with $k_{0}$ obtained from
Eq.~(\ref{eq:Bird}), collapse onto a single curve for the entire range
of (modified) Reynolds numbers, independent of the numerical values of
$\varepsilon$ and $n$. Despite the details of the model porous medium
geometry employed here as well as the complexity of the fluid
rheology, this remarkable invariance of behavior suggests that the
resulting flow properties of the system remain in the same
universality class of Newtonian fluid flow in disordered porous media.

\begin{figure}
\includegraphics[width=8cm]{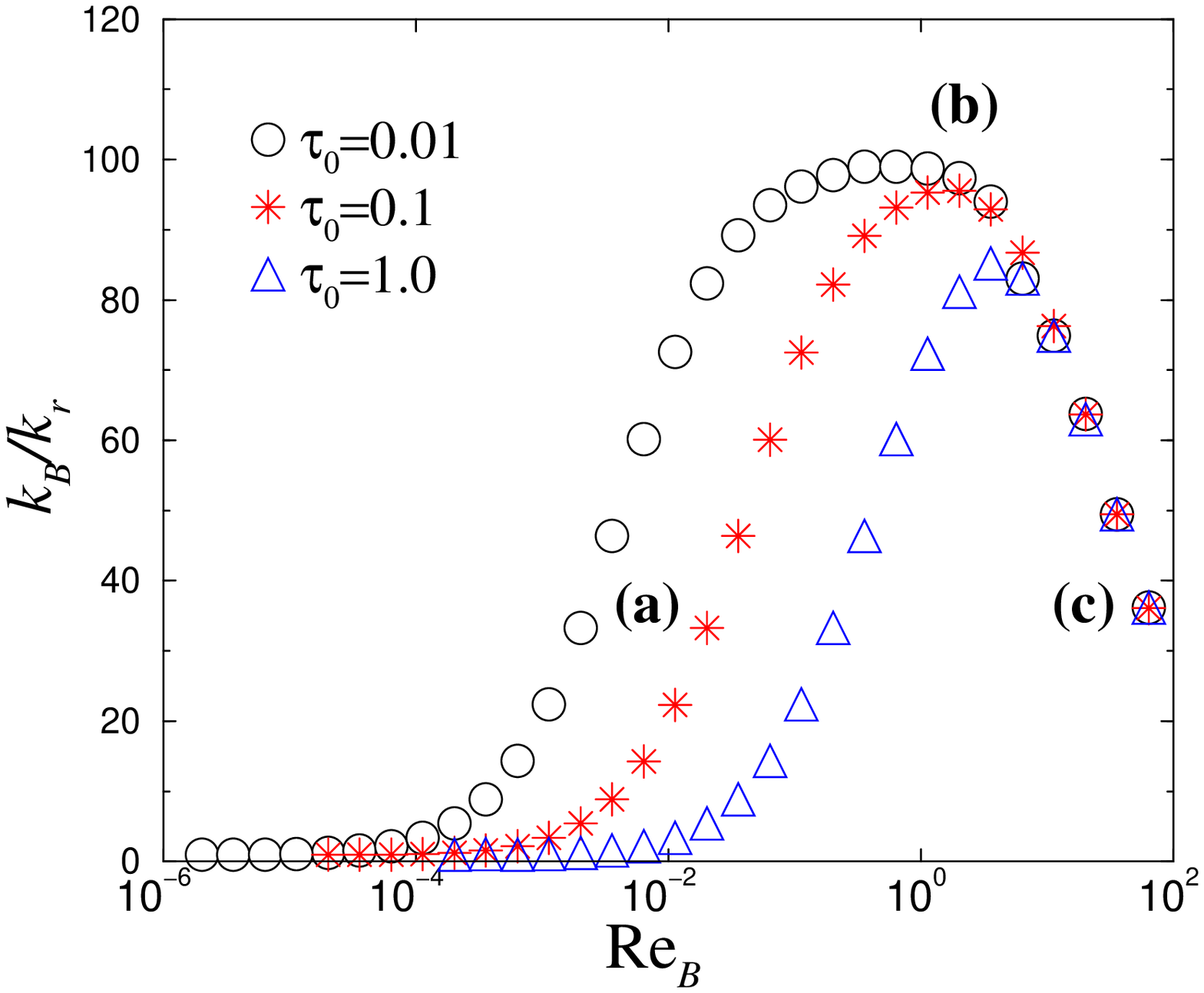}
\includegraphics[width=8cm]{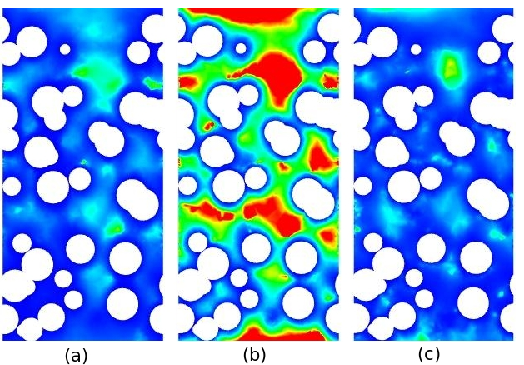}
\caption{(Color online) Flow of Bingham fluids (Herschel-Bulkley model) 
through three-dimensional porous media. The plot shows the variation
of the ratio $k_{B}/k_{r}$ with Reynolds number $\mathrm{Re}_{B}$ for
different values of the parameter $\tau_{0}$, as defined in
Eq.~(\ref{eq:Bingham}). Here $k_{r}$ corresponds to the lower limit of
$k_{B}$ at very low Reynolds conditions. The presence of maxima in all
cases is a distinctive result of the competition between rheology and
convective non-linearities. The contour plots in (a), (b) and (c) show
the spatial variation of the magnitude of the local ratio
$\left|\vec{u}\right|/\left|\nabla p\right|$ calculated on the
cross-section through the middle of the porous medium parallel to the
flow, for $\tau=0.1$ and $\mathrm{Re}_{B}= 3.5\times 10^{-2}$, $1.7$
and $35$, respectively. Their colors ranging from blue (dark) to red
(light) correspond to low and high values of
$\left|\vec{u}\right|/\left|\nabla p\right|$, respectively.}
\label{fig4}
\end{figure}

Next we present results for flow through three-dimensional porous
media of Bingham fluids with rheology given approximately by the
Herschel-Bulkley model Eq.~(\ref{eq:Bingham}). The proper way to
quantify inertial and viscous forces in this case is to define the
Reynolds number as $\mathrm{Re}_{B} \equiv \rho u_{0} d_{p}/K_{B}$. In
Fig.~\ref{fig4} we show that the linear hydraulic conductivity,
defined as $k_{B}\equiv K_{B}u_{0}L/\Delta p$, remains constant up to
a certain crossover that is proportional to the threshold $\tau_{0}$,
$\mathrm{Re}_{\times} \sim \tau_{0}$. Below this crossover, since the
fluid has Newtonian behavior with high viscosity $\mu_{0}$ everywhere
in the pore space, the flow can be macroscopically described by
Darcy's law. Above this crossover, the presence of low and high strain
rates zones in the flow leads to a nonuniform spatial distribution of
fluid viscosity, therefore increasing the overall permeability index
$k_{B}$ of the system. This behavior persists up to the point in which
inertial forces become relevant. While the specific fluid rheology
investigated here tends to enhance the flow at high $\mathrm{Re}_{B}$,
the effect of inertia is to reduce the permeability index under the
same conditions \cite{Andrade95,Andrade99}. As a result of this
competition, a maximum hydraulic conductivity can be observed at an
intermediate value of $\mathrm{Re}_{B}$ that is also dependent on the
threshold $\tau_{0}$. As shown in Fig.~\ref{fig4}, this effect is
better illustrated when we observe contour plots of the local ratio
$\left|\vec{u}\right|/\left|\nabla p\right|$ calculated at the middle
cross-section of the porous medium. To the best of our knowledge, this
condition of ``enhanced flow'' through disordered porous media
represents a novel regime of momentum transport that could have
potential applications in practical problems, e.g., chemical reactors,
chromatographic columns and switches for flow. Finally, at
sufficiently large values of $\mathrm{Re}_{B}$, the viscosity of the
fluid is uniform and therefore the local permeabilities become all the
same, regardless of the value of $\mathrm{Re}_{B}$ and $\tau_{0}$. In
this situation, all curves of $k_{B}$ collapse.

Summarizing, in spite of the non-linear nature of the fluid rheology
and the complex geometry of the interstitial pore volume, in the case
of power-law fluids, we have shown the remarkable fact that the flow
behavior can still be quantified in terms of an universal curve
extending over a broad range of Reynolds conditions and power-law
exponents. Our results for Bingham fluids are even more
striking. There the pore space geometry, fluid rheology and inertia
can combine to generate a particular condition of ``enhanced
transport'' which should be found in experiments.

We thank CNPq, CAPES, FUNCAP, FINEP, Petrobras, the National Institute of Science and Technology for Complex Systems , and the Swiss National Science Foundation (SNF) under Grante No. 116052 for financial support.

\end{document}